\UseRawInputEncoding
\documentclass[superscriptaddress, twocolumn, amsmath, amssymb, aps,prl, letterdraft, notitlepage,longbibliography]{revtex4-1}
\usepackage{graphicx,graphics,epsfig,subfigure,times,bm,bbm,amssymb,amsmath,amsfonts,amsthm,mathrsfs,MnSymbol}
\usepackage[matrix,frame,arrow]{xypic}
\usepackage[normalem]{ulem}
\usepackage{slashed}
\usepackage{color}
\usepackage[usenames,dvipsnames,svgnames,table]{xcolor}
\usepackage[english]{babel}
\usepackage{verbatim}
\usepackage{tabularx}
\definecolor{darkblue}{rgb}{0.0,0.0,0.3}
\usepackage[colorlinks=true,
            linkcolor=red,
            urlcolor= darkblue,
            citecolor=blue]{hyperref}

\newcommand{\bea}{\begin{eqnarray}}
\newcommand{\eea}{\end{eqnarray}}

\begin{document}

\title{Universal Landauer-Like Inequality from the First Law of Thermodynamics}

\author{Junjie Liu}
\email{jjliu.fd@gmail.com}
\affiliation{Department of Physics, International Center of Quantum and Molecular Structures, Shanghai University, Shanghai, 200444, China}
\author{Hanlin Nie}
\affiliation{Centre for Quantum Technologies, National University of Singapore, Block S15, 3 Science Drive 2, 117543, Singapore}

\begin{abstract}
The first law of thermodynamics, which governs energy conservation, is traditionally formulated as an equality. Surprisingly, we demonstrate that the first law alone implies a universal Landauer-like inequality linking changes in system entropy and energy. However, contrasting with the Landauer principle derived from the second law of thermodynamics, our obtained Landauer-like inequality solely relies on system information and is applicable in scenarios where implementing the Landauer principle becomes challenging. Furthermore, the Landauer-like inequality can complement the Landauer principle by establishing a dual {\it upper} bound on heat dissipation. We illustrate the practical utility of the Landauer-like inequality in dissipative quantum state preparation and quantum information erasure applications. Our findings offer new insights into identifying thermodynamic constraints relevant to the fields of quantum thermodynamics and the energetics of quantum information processing and more specifically, this approach could facilitate investigations into systems coupled to non-thermal baths or scenarios where access to bath information is limited.

\end{abstract}

\date{\today}

\maketitle

{\it Introduction.--}To ensure the sustainability of quantum technologies, conducting a comprehensive assessment of their energetic footprints is of paramount importance \cite{Auffeves.22.PRXQ}. In this context, the Landauer principle (LP) \cite{Landauer.61.IBM} marks the pioneering effort, offering an inequality that establishes a lower bound on the dissipated heat $Q(t)$ by the entropy change $\Delta S(t)$ of information-bearing degrees of freedom during the time interval $[0,t]$ (setting $\hbar=1$ and $k_B=1$):
\begin{equation}\label{eq:LP}
Q(t) \geqslant -T\Delta S(t),
\end{equation}
Here, $T$ represents the temperature of the thermal bath. Within the framework of information thermodynamics \cite{Parrondo.15.NP,Goold.16.JPA}, it is now recognized that Eq. (\ref{eq:LP}) precisely corresponds to the Clausius inequality for the total entropy production in a scenario with a single thermal bath \cite{Esposito.10.NJP,Reeb.14.NJP,Landi.21.RMP}, with $Q(t)$ and $\Delta S(t)$ identified as the averaged energy change of the thermal bath $\Delta E_B(t)$ and the system's von Neumann entropy production, respectively \cite{Esposito.10.NJP,Reeb.14.NJP,Landi.21.RMP}. Here, we define $\Delta A(t) \equiv A(t)-A(0)$ for an arbitrary quantity $A$. Nevertheless, this thermodynamic interpretation of Eq. (\ref{eq:LP}) necessitates energy measurements of a thermal bath, which are often challenging to implement due to limited control or access.

One established strategy for circumventing the bath energy measurement issue is to employ a dynamical map that converges to the system Gibbs state. A widely accepted implementation employs a quantum Lindblad master equation with Lindblad operators satisfying the detailed balance condition \cite{Alicki.87.NULL}. By adopting this approach, a Clausius inequality is obtained, where the dissipated heat $Q(t)=-\Delta E_S(t)$ is directly linked to the system's averaged energy change $\Delta E_S(t)$ which is experimentally friendly \cite{Spohn.78.JMP}. This favorable result, however, comes at the expense of requiring weak system-bath couplings \cite{Cresser.21.PRL}.

The LP has been refined \cite{Sagawa.09.PRL,Esposito.10.NJP,Hilt.11.PRE,Deffner.13.PRX,Lorenzo.15.PRL,Dago.21.PRL,Riechers.21.PRA} and generalized \cite{Goold.15.PRL,Esposito.11.EPL,Campbell.17.PRA,Reeb.14.NJP,Browne.14.PRL,Bera.17.NC,Miller.20.PRL,Proesmans.20.PRL,Saito.22.PRL,LeeJ.22.PRL,Dago.22.PRL,Timpanaro.20.PRL} in response to its fundamental and conceptual significance. Recent experimental verifications \cite{Peterson.16.PRSA,Yan.18.PRL,Gaudenzi.18.NP} have further advanced its validation. However, challenges persist when implementing the LP Eq. (\ref{eq:LP}) and its generalizations in specific scenarios. Non-thermal baths frequently encountered at the nanoscale \cite{Niedenzu.16.NJP,Bera.19.Q,Klaers.19.PRL,Macchiavello.20.PRA,Elouard.23.PRXQ} lead to an ambiguous definition of thermodynamic temperature. Additionally, calculating the total entropy production can be challenging when the final state is pure \cite{Abe.03.PRA,Santos.17.PRL}, as is the case in quantum state preparation tasks \cite{Kraus.08.PRA,Diehl.08.NP,Verstraete.09.NP}, or when systems undergo unidirectional transitions \cite{Busiello.20.PRR}, thereby limiting the universality of the LP. It is also worth noting that the existing LPs are typically derived for setups immersed in a single bath \cite{Landauer.61.IBM,Landi.21.RMP,Esposito.10.NJP,Reeb.14.NJP,Lorenzo.15.PRL,Klaers.19.PRL,Proesmans.20.PRL,Miller.20.PRL,LeeJ.22.PRL,Saito.22.PRL,Oriols.23.EPJP}, hindering their applicability in systems coupled to multiple baths. Even within the validity regime of the LP, the existence of a dual upper bound for the dissipated heat remains largely elusive.

Exerting additional refinements to overcome the limitations of the LP represents a formidable task. In this study, we leverage the equality-based first law of thermodynamics to derive universal Landauer-like inequalities for undriven and driven quantum systems [Eqs. (\ref{eq:inequality}) and (\ref{eq:inequality_ge}), respectively]. These inequalities, originating from a distinct framework, offer applicability in scenarios where the LP fails. Moreover, they can complement the LP in its validity regime by providing dual upper bounds on dissipated heat, effectively constraining it from both sides. We showcase the practical utility of these Landauer-like inequalities in a dissipative quantum Bell state preparation process \cite{LiD.18.PRA,LiuJ.23.PRA}, where the LP is inapplicable, and a quantum information erasure task on a driven qubit \cite{Miller.20.PRL,Riechers.21.PRA,Saito.22.PRL}, where the LP can be combined.

{\it Landauer-like inequality for undriven systems.--}We first consider a generic undriven quantum system with a Hamiltonian $H_S$ and a time-dependent density matrix $\rho_S(t)$, allowing for non-equilibrium conditions. Introducing a reference Gibbsian state $\rho_{th}=e^{-\beta_{R}H_S}/Z_S$, where $\beta_{R}=T_{R}^{-1}$ is an inverse reference parameter and $Z_S\equiv\mathrm{Tr}[e^{-\beta_{R}H_S}]$ is the partition function, one can obtain a universal form of the first law of thermodynamics which reduces to that in \cite{Gardas.15.PRE} for isothermal processes,
\begin{equation}\label{eq:first_law}
E_S(t)~=~T_R S(t)+\mathcal{F}(t).
\end{equation}
Here, $E_S(t)\equiv\mathrm{Tr}[H_S\rho_S(t)]$ denotes the average system energy, $S(t)=-\mathrm{Tr}[\rho_S(t)\ln \rho_S(t)]$, and $\mathcal{F}$ represents the nonequilibrium information free energy \cite{Deffner.12.A,Deffner.13.PRX,Gardas.15.PRE,Parrondo.15.NP}, $\mathcal{F}(t)\equiv F+T_{R}D[\rho_S(t)||\rho_{th}]$ where $F\equiv-T_R\ln Z_S$ and $D[\rho_1||\rho_2]\equiv\mathrm{Tr}[\rho_1(\ln\rho_1-\ln\rho_2)]$ is the relative entropy of states $\rho_{1,2}$.

From Eq. (\ref{eq:first_law}), we find that $\Delta S(t)= \beta_R\Big[\Delta E_S(t)-T_RD[\rho_S(t)||\rho_{th}]+T_RD[\rho_S(0)||\rho_{th}]\Big]$. To obtain general bounds enabling direct evaluations, we fix the reference state $\rho_{th}$ by ensuring it possesses the same entropy as the actual initial state $\rho_S(0)$ \cite{Sparaciari.17.NC,Niedenzu.19.Q,Bera.19.Q,Macchiavello.20.PRA,Elouard.23.PRXQ},
\begin{equation}\label{eq:eq_entropy}
\mathrm{Tr}[\rho_S(0)\ln\rho_S(0)]~=~\mathrm{Tr}[\rho_{th}\ln\rho_{th}].
\end{equation}
 Since $\mathrm{Tr}[\rho_{th}\ln\rho_{th}]$ is monotonically related to $\beta_{R}$ within the interval $0\le S(t)\le \ln d_S$, where $d_S$ represents the dimension of $H_S$, a unique solution for $\beta_{R}$ can be found from Eq. (\ref{eq:eq_entropy}) once $\rho_S(0)$ is specified \cite{Elouard.23.PRXQ}. It is important to emphasize that $\beta_{R}$ is a parameter determined solely by the system information. Only when the initial state is thermal, does $T_R$ acquire a meaningful thermodynamic interpretation.

With Eq. (\ref{eq:eq_entropy}), one finds $\Delta E_S(t)+T_RD(\rho_S(0)||\rho_{th})=\mathrm{Tr}\left\{\left[\rho_S(t)-\rho_{th}\right]H_S\right\}\equiv\Delta E_S^R(t)$, namely, one maps quantum relative entropy onto energy contrast with respect to the reference state. Inserting it into the expression for $\Delta S(t)$ above Eq. (\ref{eq:eq_entropy}), we attain an inequality
\begin{equation}\label{eq:inequality}
\beta_{R}\Delta E_S^R(t)-\Delta S(t)~\geqslant~0.
\end{equation}
The universality of inequality Eq. (\ref{eq:inequality}) arises from its derivation without any approximations. Noting that the left-hand-side of Eq. (\ref{eq:inequality}) denoted as $\mathcal{P}\equiv\beta_{R}\Delta E_S^R(t)-\Delta S(t)$ equals the relative entropy $D(\rho_S(t)||\rho_{th})$, the equality condition of Eq. (\ref{eq:inequality}) is satisfied when the system approaches the reference state within finite times, i.e., $\rho_S(t)=\rho_{th}$. Thus, Eq. (\ref{eq:inequality}) emerges as a consequence of the non-negative distance of $\rho_S(t)$ from an initially determined reference state $\rho_{th}$. Adopting the quantum coherence definition $\mathrm{Coh}(t)\equiv S'(t)-S(t)$ with diagonal entropy $S'(t)\equiv -\mathrm{Tr}(\Pi[\rho_S(t)]\ln\Pi[\rho_S(t)])$ and $\Pi[\rho_S(t)]\equiv\sum_n|E_n\rangle\langle E_n|\rho_S(t)|E_n\rangle\langle E_n|$ in the energy basis $\{|E_n\rangle\}$ of $H_S$ \cite{Francica.19.PRE}, we have $\Delta S(t)=\Delta S'(t)-\Delta \mathrm{Coh}(t)$ which allows us to identify the quantum coherence contribution to the bound \cite{SM}.

We term the inequality Eq. (\ref{eq:inequality}) as a Landauer-like one since it {\color{red}also} imposes constraints on energy change using system entropy production. However, the inequality Eq. (\ref{eq:inequality}) distinguishes itself from the LP Eq. (\ref{eq:LP}) in two key aspects. Firstly, Eq. (\ref{eq:inequality}) operates within a broader framework, relying solely on system information. This feature enables its applications in scenarios where the LP may fail due to limited prior knowledge of bath information or ill-defined quantities associated with the second law of thermodynamics \cite{Abe.03.PRA,Santos.17.PRL,LiuJ.23.PRA}. Moreover, unlike the LP and its generalizations \cite{Landauer.61.IBM,Landi.21.RMP,Esposito.10.NJP,Reeb.14.NJP,Lorenzo.15.PRL,Klaers.19.PRL,Proesmans.20.PRL,Miller.20.PRL,LeeJ.22.PRL,Saito.22.PRL,Oriols.23.EPJP} relying on a single bath assumption, Eq. (\ref{eq:inequality}) can be applied to systems coupled to multiple baths. 
 
Secondly, we derive the inequality Eq. (\ref{eq:inequality}) {\it exclusively} from the first law of thermodynamics Eq. (\ref{eq:first_law}). This unique foundation enables Eq. (\ref{eq:inequality}) to offer {\it complementary} constraints to the LP Eq. (\ref{eq:LP}). To clarify, we decompose $\Delta E_S^R(t)=\Delta E_S(t)+\Delta E_S^{\mathrm{in}}$, where $\Delta E_S(t)\equiv \mathrm{Tr}\left\{\left[\rho_S(t)-\rho_{S}(0)\right]H_S\right\}$ represents the actual system energy change and $\Delta E_S^{\mathrm{in}}\equiv\mathrm{Tr}\left\{\left[\rho_S(0)-\rho_{th}\right]H_S\right\}$ accounts for the possible initial energy contrast. Inserting this decomposition into Eq. (\ref{eq:inequality}), a universal upper bound on $-\Delta E_S(t)$ is obtained,
\begin{equation}\label{eq:DSP_bound}
-\Delta E_S(t)~\leqslant~\mathcal{Q}_u(t).
\end{equation}
Here, we have defined $\mathcal{Q}_u(t)\equiv \Delta E_S^{\mathrm{in}}-T_R\Delta S(t)$. Eq. (\ref{eq:DSP_bound}) estimates the amount of energy that the system {\it at most} dissipates. In arriving at the upper bound in Eq. (\ref{eq:DSP_bound}), we have assumed a non-negative reference parameter $T_R$ which can cover typical initial states \footnote{For initial states which correspond to a negative reference parameter $T_R$, we will receive instead a lower bound.}.

At weak system-bath couplings, $Q(t)=-\Delta E_S(t)$. Combining Eq. (\ref{eq:DSP_bound}) with the LP Eq. (\ref{eq:LP}), we obtain dual constraints on the dissipated heat from both sides
\begin{equation}\label{eq:both_side}
-T\Delta S(t)~\leqslant~Q(t)~\leqslant~\mathcal{Q}_u(t).
\end{equation}
Eqs. (\ref{eq:inequality})-(\ref{eq:both_side}) constitute our first main results. It is worth noting that the whole Eq. (\ref{eq:both_side}) applies to systems weakly coupled to a thermal bath at temperature $T$. Interestingly, when we set $\beta_R=\beta$ and $\rho_S(0)=\rho_{th}$, we find $\Delta E_S^{\mathrm{in}}=0$ and $\mathcal{Q}_u(t)=-T\Delta S(t)$. Consequently, the upper and lower bounds in Eq. (\ref{eq:both_side}) coincide, indicating a vanishing dissipated heat {\footnote{For a undriven system coupled a thermal bath at the temperature $T$, an initial thermal state $\propto e^{-H_S/T}$ would imply that the system is already at the thermal equilibrium in the sense that $\rho_S(t)=\rho_S(0)=\rho_{th}$. Hence the von Neumann entropy change $\Delta S(t)$ becomes zero.}}. Importantly, the left-hand-side of Eq. (\ref{eq:both_side}) is derived using the original LP Eq. (\ref{eq:LP}), and one can employ its generalizations (for instance, the one in Ref. \cite{Saito.22.PRL}) to tighten the lower bound while maintaining the validity of Eq. (\ref{eq:both_side}). 

{\it Landauer-like inequality for driven systems.--}We now generalize Eqs. (\ref{eq:inequality})-(\ref{eq:both_side}) to account for driven systems with a time-dependent Hamiltonian $H_S(t)$. By introducing an instantaneous reference Gibbsian state $\rho_{th}(t)=e^{-\beta_R(t)H_S(t)}/Z_S(t)$ with $\beta_R(t)=1/T_R(t)$ a time-dependent inverse reference parameter and $Z_S(t)=\mathrm{Tr}[e^{-\beta_R(t)H_S(t)}]$, the first law of thermodynamics still takes a similar form as Eq. (\ref{eq:first_law}), $E_S(t)=T_R(t)S(t)+\widetilde{\mathcal{F}}(t)$ but with $E_S(t)\equiv \mathrm{Tr}[H_S(t)\rho_S(t)]$ and $\widetilde{\mathcal{F}}(t)\equiv F(t)+T_R(t)D[\rho_S(t)||\rho_{th}(t)]$; $F(t)=-T_R(t)\ln Z_S(t)$. The change in the system's von Neumann entropy can be expressed as $\Delta S(t)=\beta_R(t)[E_S(t)-\widetilde{\mathcal{F}}(t)]-\beta_R(0)[E_S(0)-\widetilde{\mathcal{F}}(0)]$.

In this scenario, we fix the instantaneous reference state by requiring an instantaneous equivalence of entropy, allowing for a unique solution for $\beta_R(t)$ \cite{Elouard.23.PRXQ}
\begin{equation}
\mathrm{Tr}[\rho_S(t)\ln\rho_S(t)]~=~\mathrm{Tr}[\rho_{th}(t)\ln\rho_{th}(t)].
\end{equation}
By exploiting the above condition at $t=0$, we can transfer $D[\rho_S(0)||\rho_{th}(0)]=\mathrm{Tr}\left\{\left[\rho_{th}(0)-\rho_S(0)\right]\ln\rho_{th}(0)\right\}=-\beta_R(0)\mathrm{Tr}\left\{\left[\rho_{th}(0)-\rho_S(0)\right]H_S(0)\right\}$. Inserting this expression into the above equation for $\Delta S(t)$ and rearranging terms, we obtain a generalized Landauer-like inequality that extends Eq. (\ref{eq:inequality}),
\begin{equation}\label{eq:inequality_ge}
\beta_R(0)\Delta\widetilde{E}_S^R(t)-\Delta S(t)+\mathcal{C}(t)~\geqslant~0.
\end{equation}
Here, we define $\Delta\widetilde{E}_S^R(t)\equiv E_S(t)-E_S^{th}(0)$ with $E_S^{th}(0)=\mathrm{Tr}[H_S(0)\rho_{th}(0)]$, $\mathcal{C}(t)\equiv \Delta \beta_R(t)E_S(t)+\ln\frac{Z_S(t)}{Z_S(0)}$ with $\Delta \beta_R(t)=\beta_R(t)-\beta_{R}(0)$. $\mathcal{C}(t)$ vanishes when $H_S$ becomes time independent, as then $\beta_R$ has no time dependence. The equality condition of Eq. (\ref{eq:inequality_ge}) is satisfied when $\rho_S(t)=\rho_{th}(t)$, as the left-hand-side of Eq. (\ref{eq:inequality_ge}) precisely corresponds to $D[\rho_S(t)||\rho_{th}(t)]$. Similar to Eq. (\ref{eq:inequality}), one can still decompose $\Delta S(t)=\Delta S'(t)-\Delta \mathrm{Coh}(t)$ with $S'(t)$ expressed in terms of the instantaneous energy basis \cite{Francica.19.PRE} and identify the quantum coherence contribution to the bound \cite{SM}.

Eq. (\ref{eq:inequality_ge}) leads to generalizations of Eqs. (\ref{eq:DSP_bound}) and (\ref{eq:both_side}), forming the second main results of this study,
\bea
&&-\Delta E_S(t)~\leqslant~\widetilde{\mathcal{Q}}_u(t),\label{eq:bounds_ge}\\
&&-T\Delta S(t)~\leqslant~Q(t)~\leqslant~\widetilde{\mathcal{Q}}_u(t)+W(t).\label{eq:both_side_ge}
\eea
Here, $\Delta E_S(t)=\mathrm{Tr}[\rho_S(t)H_S(t)-\rho_S(0)H_S(0)]$. We define $\widetilde{\mathcal{Q}}_u(t)\equiv \Delta\widetilde{E}_S^{in}-T_R(0)\Delta S(t)+T_R(0)\mathcal{C}(t)$ with $\Delta\widetilde{E}_S^{in}\equiv \mathrm{Tr}\left\{\left[\rho_S(0)-\rho_{th}(0)\right]H_S(0)\right\}$ and $W(t)=\int_0^t\mathrm{Tr}[\dot{H}_S(t')\rho_S(t')]dt'$ denotes the work performed on the system with $\dot{A}(t)\equiv dA(t)/dt$ for an arbitrary $A(t)$. We remark that Eq. (\ref{eq:bounds_ge}) holds under the assumption of a non-negative initial reference parameter $T_R(0)$, and Eq. (\ref{eq:both_side_ge}) applies to driven systems weakly coupled to a thermal bath at the temperature $T$. The usual LP holds in driven systems with dissipated heat $Q(t)=-\int_0^t\mathrm{Tr}[H_S(t')\dot{\rho}_S(t')]dt'$ \cite{Saito.22.PRL} and $\Delta E_S(t)=-Q(t)+W(t)$. Eqs. (\ref{eq:DSP_bound}) and (\ref{eq:both_side}) are recovered when $H_S$ becomes time independent. Notably, the upper and lower bounds in Eq. (\ref{eq:both_side_ge}) differ, even when $\rho_S(0)=\rho_{th}(0)$ and $T_R(0)=T$, as the upper bound includes a nonzero work contribution.

{\it Application 1: Dissipative quantum state preparation.--}To validate the framework, we first consider the dissipative quantum state preparation (DQSP) which harnesses dissipative processes to prepare useful quantum pure states. One generally utilizes a Markovian quantum Lindblad master equation to this end \cite{Plenio.99.PRA,Kraus.08.PRA,Diehl.08.NP,Verstraete.09.NP},
\begin{equation}\label{eq:lindblad}
 \frac{d}{dt}\rho_S(t)~=~-i[H_S,\rho_S(t)]+\sum_{\mu=1}\gamma_{\mu}\mathcal{D}[L_{\mu}]\rho_S(t) 
 \end{equation}
Here, $\gamma_{\mu}\geqslant 0$ is the damping coefficient of channel $\mu$, $\mathcal{D}[L_{\mu}]\rho=L_{\mu}\rho L_{\mu}^{\dagger}-\frac{1}{2}\{L_{\mu}^{\dagger}L_{\mu},\rho\}$ is the Lindblad superoperator with $L_{\mu}$ the Lindblad jump operator and $\{A,B\}=AB+BA$. The final stationary state would be a pure one
$|\Phi\rangle$ when it fulfills the conditions $H_S|\Phi\rangle=E_{n}|\Phi\rangle$ and $L_{\mu}|\Phi\rangle=0,~\forall\mu$ \cite{Kraus.08.PRA}.

The DQSP process is typically accomplished through dissipation engineering \cite{Harrington.22.NRP} involving non-thermal baths \cite{Leghtas.15.S}, wherein the engineered Lindblad jump operators do not adhere to the detailed balance condition \cite{Kraus.08.PRA}. Furthermore, the final pure state challenges the conventional definition of total entropy production, rendering it ill-defined \cite{Abe.03.PRA,Santos.17.PRL}. As a result, the LP Eq. (\ref{eq:LP}) cannot quantify the thermodynamic cost of DQSP processes \cite{LiuJ.23.PRA}. In contrast, the inequality Eq. (\ref{eq:inequality}) remains applicable.
%
\begin{figure}[thb!]
 \centering
\includegraphics[width=1\columnwidth]{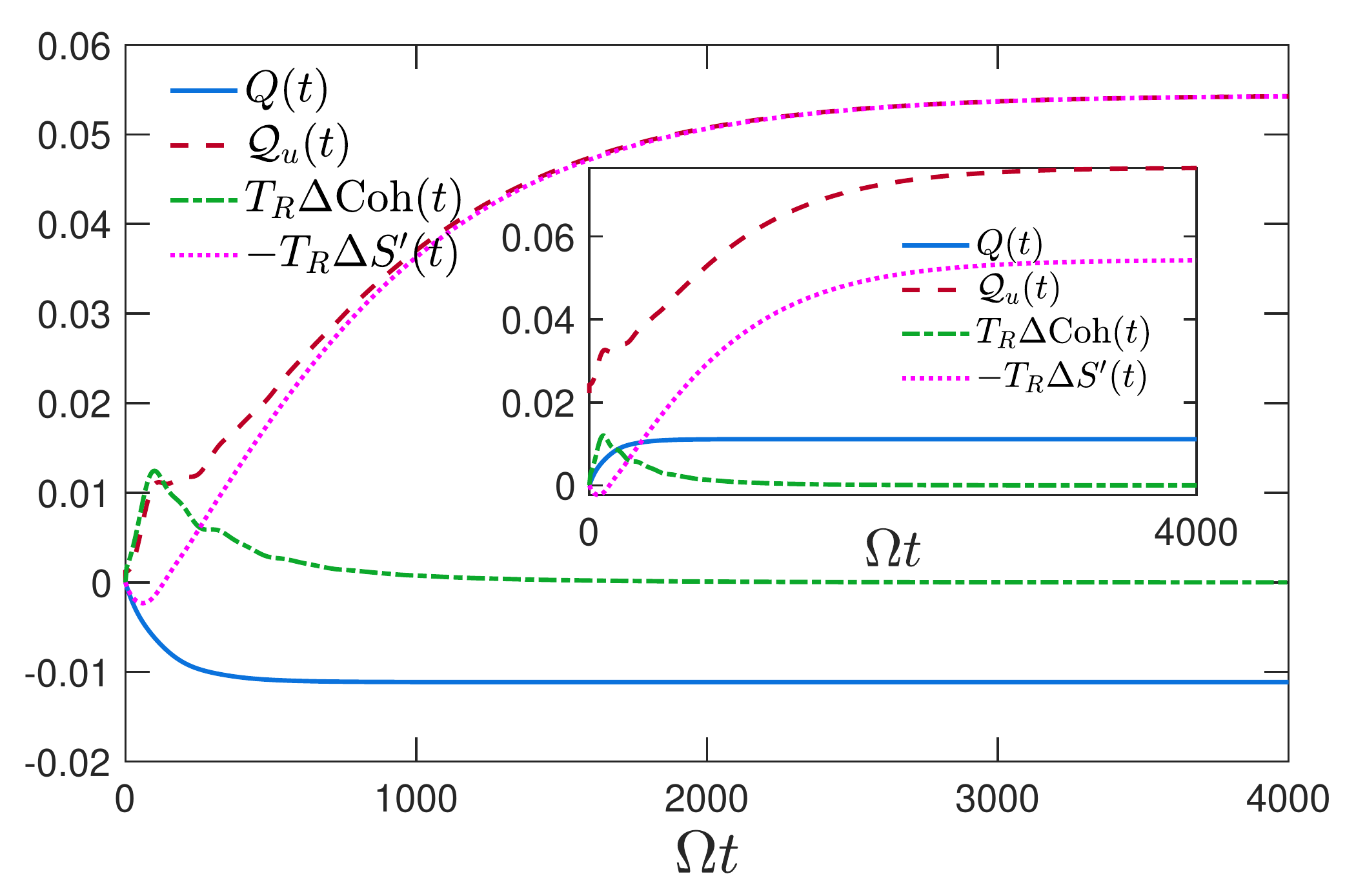} 
\caption{Results for the time-dependent dissipated heat $Q(t)$ (blue solid line), the upper bound $\mathcal{Q}_u(t)$ given by Eq. (\ref{eq:both_side}) (red dashed line), quantum coherence contribution $T_R\Delta \mathrm{Coh}(t)$ (green dashed-dotted line) and diagonal entropy contribution $-T_R\Delta S'(t)$ (magenta dotted line). Main: $\rho_S(0)=\rho_{th}$ with $\beta_R=30$. Inset: $\rho_S(0)$ has the same diagonal elements with $\rho_{th}$ but sorted in an increasing order with respect to an ordered energy basis with increasing eigen-energies. We set $\Omega=2\pi$ MHz as the unit and adopt experimental values $(\Omega_2,\omega,\gamma)=2\pi\times(0.02,0.01,0.03)$ MHz \cite{Grankin.14.NJP}. 
}
\protect\label{fig:DSP}
\end{figure}

In this application, we validate the upper bound in Eq. (\ref{eq:both_side}); Noting $Q(t)=-\Delta E_S(t)$ for the description Eq. (\ref{eq:lindblad}) \cite{Spohn.78.JMP,Saito.22.PRL}. We consider a concrete DQSP setup proposed by Ref. \cite{LiD.18.PRA} consisting of two $\Lambda$-type three-level Rydberg atoms, each one contains two ground states $|0\rangle$ and $|1\rangle$, and one Rydberg state $|r\rangle$. Combining an unconventional Rydberg pumping mechanism with the spontaneous emission of two atoms, Ref. \cite{LiD.18.PRA} showed that one can dissipatively generate the Bell state $|\Phi\rangle=(|00\rangle-|11\rangle)/\sqrt{2}$ with $|00(11)\rangle$ being understood as $|0(1)\rangle\otimes|0(1)\rangle$. Elements in Eq. (\ref{eq:lindblad}) read \cite{LiD.18.PRA}: $H_S=\Omega_2(|10\rangle\langle r0|+|01\rangle\langle 0r|)+\omega\Big[(|11\rangle+|00\rangle)\otimes (\langle 01|+\langle 10|)\Big]+\mathrm{H.c.}$ ($\mathrm{H.c.}$ denotes Hermitian conjugate), $\gamma_{\mu}=\gamma/2$, and four Lindblad jump operators describing spontaneous emission $L_1=|01\rangle\langle 0r|$, $L_2=|00\rangle\langle 0r|$, $L_3=|10\rangle\langle r0|$ and $L_4=|00\rangle\langle r0|$. It is evident that the Bell state satisfies conditions $H_S|\Phi\rangle=0$ and $L_{1,2,3,4}|\Phi\rangle=0$ as required by the DQSP scheme. Nevertheless, one should bear in mind that the adopted model overlooks other decaying channels such as the dephasing one.

In Fig. \ref{fig:DSP}, we depict a set of results for both $Q(t)$ and its upper bound $\mathcal{Q}_u(t)$ for the aforementioned model. In the main plot, we take $\rho_S(0)=\rho_{th}$ with $\beta_R=30$, while in the inset, we consider a non-thermal-form initial state obtained by sorting the same diagonal elements of $\rho_{th}$ in an increasing order with respect to an ordered energy basis with increasing eigen-energies. As can be seen from the figure, $\mathcal{Q}_u(t)$ indeed bounds the dissipated heat $Q(t)$ from above. We also note that the contribution from quantum coherence, $T_R\Delta \mathrm{Coh}(t)$, is only impactful at short times and gradually diminishes as time progresses. Consequently, the dominant factor governing the upper bound at extended times is the reduction in diagonal entropy, $-T_R\Delta S'(t)$. The finite distance $\mathcal{Q}_u(t)-Q(t)=T_R\mathcal{P}$ [See definition below Eq. (\ref{eq:inequality})] becomes maximum at long times as $|\Phi\rangle$ significantly deviates from a full-rank reference state. One can potentially reduce the distance by strategically adjusting the initial conditions to mitigate the contribution $-T_R\Delta S'(t)$ \footnote{The magnitude of the initial energy contrast $\Delta E_S^{\mathrm{in}}$ with varying initial states can maintain at a low level by noting that the initial condition chose for the inset of Fig. 1 defines the largest initial energy contrast $\Delta E_S^{\mathrm{in}}$ among full-rank initial states with the same diagonal elements.}. 
Notably, the initial state $\rho_{th}$ (more generally, passive states \cite{LiuJ.23.PRA}) leads to a negative dissipated heat as can be seen from Fig. \ref{fig:DSP}; This occurs because the final Bell state is the third excited state of the system and the system gains energy from the environment to complete the preparation process  \cite{LiuJ.23.PRA}. With the LP Eq. (\ref{eq:LP}), one can just deduce $\beta Q(t)\ge-\Delta S(t)\ge 0$, given that $\Delta S(t)\le 0$ in simulations. However, due to the undefined nature of $\beta$, extracting information about $Q(t)$ alone is impossible.

{\it Application 2: Information erasure.--}We then turn to an information erasure model in which a driven qubit is coupled to a thermal bath at the temperature $T=\beta^{-1}$ \cite{Miller.20.PRL,Riechers.21.PRA,Saito.22.PRL}. The system is described by a time-dependent Hamiltonian
\begin{equation}\label{eq:hst}
H_S(t)~=~\frac{\varepsilon(t)}{2}\Big(\cos[\theta(t)]\sigma_z+\sin[\theta(t)]\sigma_x\Big).
\end{equation}
Here, $\sigma_{x,z}$ are the Pauli matrices, $\varepsilon(t)$ and $\theta(t)$ are time-dependent control parameters. We adopt the control protocols $\varepsilon(t)=\varepsilon_0+(\varepsilon_{\tau}-\varepsilon_0)\sin(\pi t/2\tau)^2$ and $\theta(t)=\pi(t/\tau-1)$ \cite{Miller.20.PRL}. The evolution of $\rho_S(t)$ is still governed by the quantum Lindblad master equation Eq. (\ref{eq:lindblad}) but with the time-dependent Hamiltonian in Eq. (\ref{eq:hst}) and two time-dependent jump operators $L_1(t)=\sqrt{\varepsilon(t)[N_B(t)+1]}|0_t\rangle\langle 1_t|$, $L_2(t)=\sqrt{\varepsilon(t)N_B(t)}|1_t\rangle \langle 0_t|$ \cite{Saito.22.PRL}. Here, $|0_t\rangle$ ($|1_t\rangle$) is the instantaneous ground (excited) state of $H_S(t)$, $N_B(t)=1/(e^{\beta\varepsilon(t)}-1)$, and $\gamma_{1,2}=\gamma$.
%
\begin{figure}[thb!]
 \centering
\includegraphics[width=1\columnwidth]{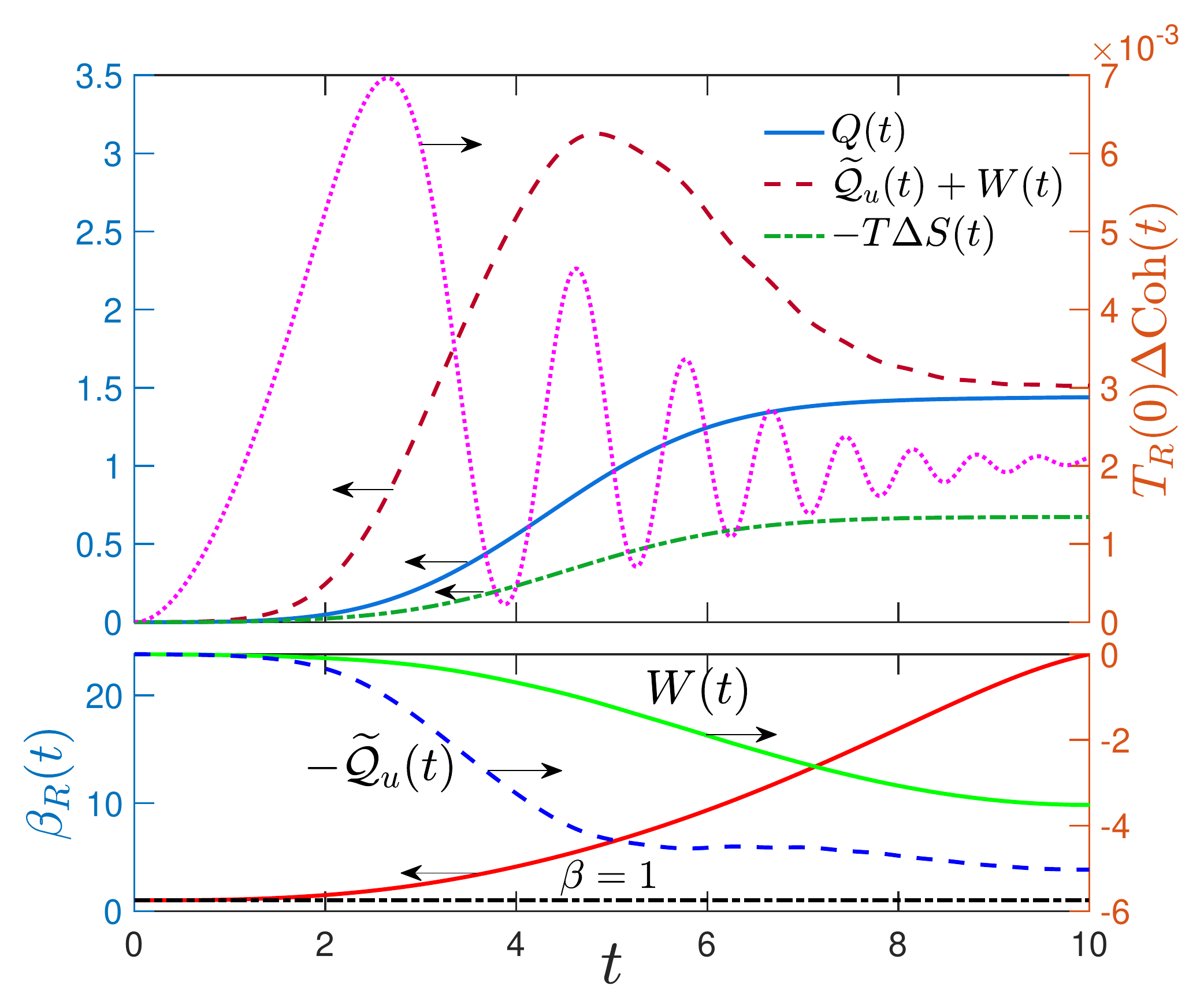} 
\caption{Validating Eq. (\ref{eq:both_side_ge}) with an initial thermal state $\rho_S(0)=e^{-\beta H_S(0)}/\mathrm{Tr}[e^{-\beta H_S(0)}]$. Upper panel: The dissipated heat $Q(t)$ (blue solid line), derived upper bound $\widetilde{\mathcal{Q}}_u(t)+W(t)$ (red dashed line), the LP bound $-T\Delta S(t)$ (green dotted-dashed line) and quantum coherence contribution $T_R(0)\Delta\mathrm{Coh}(t)$ (right axis, magenta dotted line). Lower panel: The time-dependent inverse reference temperature $\beta_R(t)$ (left axis), the work $W(t)$ and $-\widetilde{\mathcal{Q}}_u(t)$ (right axis). Parameters are $\gamma=0.2$, $\beta=T^{-1}=1$, $\varepsilon_0=0.4$, $\varepsilon_{\tau}=10$ and $\tau=10$.
}
\protect\label{fig:erasure}
\end{figure}

For this driven setup, we consider validating Eq. (\ref{eq:both_side_ge}). In the case of a maximally mixed initial state $\rho_S(0)=\mathrm{I}/2$ with `$\mathrm{I}$' a $2\times 2$ identity matrix \cite{Saito.22.PRL}, the upper bound in Eq. (\ref{eq:both_side_ge}) is trivially divergent due to the divergent initial reference parameter. For demonstration, we consider an easily-prepared initial thermal state $\rho_S(0)=e^{-\beta H_S(0)}/\mathrm{Tr}[e^{-\beta H_S(0)}]$ \footnote{One can obtain an initial thermal state by coupling the system with the Hamiltonian $H_S(0)$ to the thermal bath for sufficiently long time without activating the driving fields.}. In this case, the upper bound in Eq. (\ref{eq:both_side_ge}) becomes finite and nontrivial. A set of numerical results are depicted in Fig. \ref{fig:erasure}. 

In the upper panel of Fig. \ref{fig:erasure}, we observe that the dissipated heat $Q(t)$ is indeed bounded by both the derived upper bound $\widetilde{\mathcal{Q}}_u(t)+W(t)$ and the LP lower bound $-T\Delta S(t)$. Opting for $\beta_R(0)=\beta$, both bounds in Eq. (\ref{eq:both_side_ge}) incorporate an identical quantum coherence term $T_R(0)\Delta\mathrm{Coh}(t)$. This term exhibits oscillations but does not significantly contribute to the bound. However, decreasing $\tau$ amplifies the oscillation magnitude of $T_R(0)\Delta\mathrm{Coh}(t)$; see \cite{SM} for more details. In the lower panel of Fig. \ref{fig:erasure}, we notice the monotonic increase of the inverse reference parameter $\beta_R(t)$ from the initial actual inverse temperature $\beta_R(0)=\beta=1$ (black dotted-dashed line). This behavior allows us to utilize $\beta_R(t)$ as a monitoring tool for the effectiveness of the erasure process, driving the qubit towards the ground state with a divergent effective inverse temperature. Interestingly, we also observe a negative $W(t)$ in the lower panel of Fig. \ref{fig:erasure}, implying a work output from the driven qubit. Given analogous heat engine setups involving a single bath and two driving fields \cite{Schmiedl.08.EPL,Brandner.15.PRX,Brandner.16.PRE}, there's a compelling question about adapting the information erasure model for heat engine design--a topic for future exploration.

{\it Discussion and conclusion.--} Our Landauer-like inequalities are experimentally feasible. By utilizing quantum state tomography, the reduced system state and all relevant quantities in the inequalities, including the reference temperature, can be determined. While we focused on applications related to dissipated heat, these inequalities possess a broader scope. For instance, they can be applied to investigate irreversibility in thermal relaxation processes, especially those where entropy production plays a significant role \cite{Hasegawa.21.PRL}.

Ref. \cite{Esposito.11.EPL} revealed a nonequilibrium Landauer principle (NLP) using the non-negativity of quantum relative entropy as well. While the NLP gives rise to inequalities resembling Eqs. (\ref{eq:inequality}) and (\ref{eq:inequality_ge}), a meticulous analysis confirms their distinction from Eqs. (\ref{eq:inequality}) and (\ref{eq:inequality_ge}), with our results being more general; detailed insights are available in \cite{SM}.

In conclusion, we derived Landauer-like inequalities from the first law of thermodynamics. These inequalities complement the Landauer principle and provides new insights into identifying thermodynamic constraints. We expect our results of interests to both the fields of quantum thermodynamics and quantum information science. 

{\it Acknowledgement.--}
J. Liu acknowledges supports from Shanghai Pujiang Program (Grant No. 22PJ1403900), the National Natural Science Foundation of China (Grant No. 12205179), and start-up funding of Shanghai University. H. Nie is supported by CQT PhD programme.

%

\newpage
\renewcommand{\thesection}{\Roman{section}} 
\renewcommand{\thesubsection}{\Alph{subsection}}
\renewcommand{\theequation}{S\arabic{equation}}
\renewcommand{\thefigure}{S\arabic{figure}}
\renewcommand{\thetable}{S\arabic{table}}
\setcounter{equation}{0}  
\setcounter{figure}{0}

\begin{widetext}

{\Large{\bf Supplemental Material:} Universal Landauer-Like Inequality from the First Law of Thermodynamics}

\renewcommand{\theequation}{S\arabic{equation}}
\renewcommand{\thefigure}{S\arabic{figure}}
\setcounter{equation}{0}  

In Sec. I of this supplemental material, we first derive inequalities from a nonequilibrium Landauer principle \cite{Esposito.11.EPL} and contrast them with our inequalities so as to show that ours are independent of them and more general. Then in Sec. II, we rewrite our inequalities to explicitly highlight the contributions of quantum coherence and depict additional numerical results for the quantum information erasure model. Just as in the main text we work in units where $\hbar=1$ and $k_B=1$.

\section{I. Bounds from a nonequilibrium Landaure principle and comparisons}\label{sec:1}
In Ref. \cite{Esposito.11.EPL}, a nonequilibrium Landauer principle has been derived for driven systems with Hamiltonian $H_S(t)$ coupled to a thermal bath at a temperature $T\equiv\beta^{-1}$:
\begin{equation}\label{eq:s1}
F(t)~\ge~F^{\mathrm{eq}}(t).
\end{equation}
Here, $F(t)=E_S(t)-TS(t)$ is a nonequilibrium free energy with $E_S(t)=\mathrm{Tr}[\rho_S(t)H_S(t)]$ the system internal energy and $S(t)=-\mathrm{Tr}[\rho_S(t)\ln\rho_S(t)]$ the system von-Neumann entropy with respect to the system reduced density matrix $\rho_S(t)$, $F^{\mathrm{eq}}(t)=E_S^{\mathrm{eq}}(t)-TS^{\mathrm{eq}}(t)$ is the corresponding equilibrium free energy with $E_S^{\mathrm{eq}}(t)$ and $S^{\mathrm{eq}}(t)$ obtained by replacing $\rho_S(t)$ in $F(t)$ and $S(t)$ with an instantaneous thermal state $\rho^{\mathrm{eq}}(t)\equiv\exp[-\beta(H_S(t)-F^{\mathrm{eq}}(t))]$, respectively. For undriven systems, Eq. (\ref{eq:s1}) simply reduces to $F(t)\ge F^{\mathrm{eq}}$ where the equilibrium free energy is time-independent.

We note if one abandons the use of a thermodynamic meaningful temperature and considers generally the nonequilibrium free energy defined in Eq. (1) of the main text, $\mathcal{F}(t)=E_S(t)-T_RS(t)$, one can generalize the nonequilibrium Landauer principle Eq. (\ref{eq:s1}) to account for arbitrary processes including the isothermal one considered by Ref. \cite{Esposito.11.EPL}, \begin{eqnarray}\label{eq:s2}
        F(t)-F_{th} &=& \mathrm{Tr}[\rho(t)H_S]+T_R\mathrm{Tr}[\rho(t)\ln\rho(t)]-F_{th}\nonumber\\
        &=& -T_R\mathrm{Tr}[\rho(t)\ln\rho_{th}]+F_{th}+T_R\mathrm{Tr}[\rho(t)\ln\rho(t)]-F_{th}\nonumber\\
        &=& T_RD[\rho(t)||\rho_{th}]~\ge~0.
\end{eqnarray}
In arriving at the second line, we have utilized the relation $\ln\rho_{th}=-\beta_R(H_S-F_{th})$ for the reference thermal state $\rho_{th}$ with $T_R$ playing a role as a temperature. 

Nevertheless, one should bear in mind that the general Eq. (\ref{eq:s2}) is useless in practice until the reference state $\rho_{th}$ (or the parameter $T_R$) is specified in some ways. Otherwise, direct evaluations of quantities in Eq. (\ref{eq:s2}) are impossible. In Ref. \cite{Esposito.11.EPL}, the authors chose to specify the reference state by coupling the system to a thermal bath at a temperature $T$, thereby fixing $\rho_{th}=\rho^{\mathrm{eq}}$ and $T_R=T$. Whereas in our treatment, we instead consider fixing the reference state and parameter $T_R$ using just the system entropy which is always well-defined and measurable. By doing so, we can avoid invoking the thermal bath assumption and preserve the applicability of bounds in scenarios with non-thermal baths or even without access to bath information. This key difference in the way of fixing the reference state underpins the distinctions between bounds derived from Ref. \cite{Esposito.11.EPL} and those obtained in the main text as will be seen later.

\subsection{A.~Deriving bounds based on Ref. \cite{Esposito.11.EPL}}
In this subsection, we utilize the original nonequilibrium Landauer principle Eq. (\ref{eq:s1}) \cite{Esposito.11.EPL} to derive bounds relating system energy and entropy changes. Directly from Eq. (\ref{eq:s1}) and noting the nonequilibrium free energy defined in Ref. \cite{Esposito.11.EPL}, we find
    \begin{equation}\label{eq:21}
        F(t)-F^{\mathrm{eq}}(t)~=~E_S(t)-TS(t)-E_S^{\mathrm{eq}}(t)+TS^{\mathrm{eq}}(t)~\ge~0.
    \end{equation}
    If one defines
    \begin{eqnarray}\label{eq:22}
        \Delta E_S^{'}(t) &\equiv& E_S(t)-E_S^{\mathrm{eq}}(t),\nonumber\\
        \Delta S^{'}(t) &\equiv& S(t)-S^{\mathrm{eq}}(t),
    \end{eqnarray}
    one receive the following inequality 
    \begin{equation}\label{eq:23}
        \beta \Delta E_S^{'}(t)-\Delta S^{'}(t)~\ge~0.
    \end{equation}
    If we further introduce the following contrasts 
    \begin{eqnarray}
        \Delta E_S^{''}(t) &\equiv& E_S(t)-E_S^{\mathrm{eq}}(0),\nonumber\\
        \Delta S^{''}(t) &\equiv& S(t)-S^{\mathrm{eq}}(0),
    \end{eqnarray}
    we can rewrite inequality Eq. (\ref{eq:23}) as
    \begin{equation}\label{eq:25}
       \beta \Delta E_S^{''}(t)-\Delta S^{''}(t)+\ln\frac{Z_S^{\mathrm{eq}}(t)}{Z_S^{\mathrm{eq}}(0)}~\ge~0.
    \end{equation}
    Note that the above Eq. (\ref{eq:25}) applies to driven systems. For undriven systems, Eq. (\ref{eq:25}) reduces to
    \begin{equation}\label{eq:26}
        \beta \Delta E_S^{'''}(t)-\Delta S^{'''}(t) ~\ge~ 0.
    \end{equation}
    Here, the quantity contrasts take the forms
    \begin{eqnarray}\label{eq:27}
        \Delta E_S^{'''}(t) &\equiv& E_S(t)-E_S^{\mathrm{eq}},\nonumber\\
        \Delta S^{'''}(t) &\equiv& S(t)-S^{\mathrm{eq}}.
    \end{eqnarray}

\subsection{B.~Comparisons}
In this subsection, we contrast the above Eqs. (\ref{eq:25}) and (\ref{eq:26}) with our bounds derived in the main text. 

\subsubsection{1.~Undriven systems}
For undriven systems, we compare Eq. (\ref{eq:26}) above with Eq. (4) of the main text which reads
\begin{equation}\label{eq:s10}
    \beta_R\Delta E_S^R-\Delta S(t)~\ge~0.
\end{equation}
Here, we defined 
\begin{eqnarray}\label{eq:s11}
    \Delta E_S^R(t) &=& E_S(t)-E_S^{th},\nonumber\\
    \Delta S(t) &=& S(t)-S(0).
\end{eqnarray}
Comparing Eq. (\ref{eq:27}) with Eq. (\ref{eq:s11}), the only way to have $\beta_R=\beta$, $\Delta E_S^R(t)=\Delta E_S^{'''}(t)$ and $\Delta S(t)=\Delta S^{'''}(t)$ and thus build an equivalence between Eq. (\ref{eq:26}) and Eq. (\ref{eq:s10}) is to set $\rho(0)=\rho_{th}$ and $\rho_{th}=\rho^{\mathrm{eq}}$: The former condition means that the system's evolution starts from the reference Gibbsian state $\rho_{th}$, and the latter condition states that this reference Gibbsian state is in fact a physical thermal one $\rho^{\mathrm{eq}}$. To meet the conditions, we should assume that the system is coupled to a thermal bath and reaches the thermal equilibrium state at $t=0$. Except this special scenario, we generally have $\rho(0)\neq\rho_{th}$ (for instance, $\rho(0)$ can be a non-Gibbsian state) and thus $\Delta S(t)\neq\Delta S^{'''}(t)$, implying that Eq. (\ref{eq:s10}) is no longer equivalent to Eq. (\ref{eq:26}) derived from Ref. \cite{Esposito.11.EPL}. 

Furthermore, since $T_R$ is solely determined by the system entropy, our bound Eq. (\ref{eq:s10}) operates within a broader framework compared with Eq. (\ref{eq:26}). For instance, Eq. (\ref{eq:s10}) can be applied to systems coupled to non-thermal baths whose temperatures are ill-defined. Hence, Eq. (\ref{eq:26}) from Ref. \cite{Esposito.11.EPL} should be regarded as a special case of our bound Eq. (\ref{eq:s10}) when the system is explicitly coupled to a thermal bath and reaches the thermal equilibrium state initially.

\subsubsection{2.~Driven systems:}
For driven systems, we compare Eq. (\ref{eq:25}) above with Eq. (8) of the main text which reads
\begin{equation}\label{eq:s12}
    \beta_R(0)\Delta \widetilde{E}_S^R(t)-\Delta S(t)+\mathcal{C}(t)~\ge~0.
\end{equation}
Here, $\Delta \widetilde{E}_S^R(t)=E_S(t)-E_S^{th}(0)$ and $\mathcal{C}(t)=[\beta_R(t)-\beta_R(0)]E_S(t)+\ln\frac{Z_S^{th}(t)}{Z_S^{th}(0)}$. We first remark that for driven systems the forms of nonequilibrium free energy utilized in Ref. \cite{Esposito.11.EPL} and our study become distinct: Ref. \cite{Esposito.11.EPL} considered $F(t)=E_S(t)-TS(t)$ with a fixed thermodynamic temperature, whereas ours reads $\mathcal{F}(t)=E_S(t)-T_R(t)S(t)$ with a time-dependent reference parameter $T_R(t)$ as the system entropy can change during non-unitary processes. 

Let us first consider a special case in which $\rho(0)=\rho_{th}(0)=\rho^{\mathrm{eq}}$ and $\beta_R(0)=\beta$, we then have $\Delta \widetilde{E}_S^R(t)=\Delta E_S^{''}(t)$ and $\Delta S(t)=\Delta S^{''}(t)$. However, one still finds that $\mathcal{C}(t)\neq \ln\frac{Z_S^{\mathrm{eq}}(t)}{Z_S^{\mathrm{eq}}(0)}$ as $\beta_R(t)\neq\beta_R(0)$ at later times due to non-unitary evolutions. Beyond this special case, we generally expect $\Delta \widetilde{E}_S^R(t)\neq \Delta E_S^{''}(t)$ and $\Delta S(t)\neq \Delta S^{''}(t)$, and $\mathcal{C}(t)\neq \ln\frac{Z_S^{\mathrm{eq}}(t)}{Z_S^{\mathrm{eq}}(0)}$ persists. Hence, we remark that for driven systems our bound and that obtained based on Ref. \cite{Esposito.11.EPL} are independent. And our bound does not necessarily require the system to couple to a thermal bath.

\section{II. Quantum coherence contribution and additional numerical results}
To identify contribution of quantum coherence in inequalities derived in the main text, we follow a definition of quantum coherence in Ref. \cite{Francica.19.PRE},
\begin{equation}\label{eq:s13}
    \mathrm{Coh}(t)~\equiv~S'(t)-S(t).
\end{equation}
Here, $S'(t)\equiv -\mathrm{Tr}(\Pi_t[\rho_S(t)]\ln\Pi_t[\rho_S(t)])$ denotes a diagonal entropy with respect to a diagonal density matrix $\Pi_t[\rho_S(t)]$ under the action of a dephasing map $\Pi_t$
\begin{equation}
    \Pi_t[\rho_S(t)]\equiv\sum_n|E_n(t)\rangle\langle E_n(t)|\rho_S(t)|E_n(t)\rangle\langle E_n(t)|
\end{equation}
with $\{|E_n(t)\}$ the instantaneous energy basis of $H_S(t)$. For undriven systems, the dephasing map above becomes time-independent as the energy basis is fixed to $\{|E_n\rangle\}$. 

With the definition in Eq. (\ref{eq:s13}), we have the following decomposition of the system entropy change
\begin{equation}
    \Delta S(t)~=~\Delta S'(t)-\Delta \mathrm{Coh}(t).
\end{equation}
Hence we can rewrite the inequalities in the main text to idenfity the contribution of quantum coherence. For undriven systems, Eq. (4) of the main text can be expressed as
\begin{equation}
    \beta_R\Delta E_S^R(t)-\Delta S'(t)+\Delta \mathrm{Coh}(t)~\ge~0,
\end{equation}
and 
\begin{equation}
    \mathcal{Q}_u(t)~=~\Delta E_S^{\mathrm{in}}-T_R\Delta S'(t)+T_R\Delta \mathrm{Coh}(t)
\end{equation}
in Eq. (5) of the main text. 

For driven system, we similarly rewrite Eq. (8) of the main text as
\begin{equation}
   \beta_R(0)\Delta\widetilde{E}_S^R(t)-\Delta S'(t)+\Delta \mathrm{Coh}(t)+\mathcal{C}(t)~\ge~0,
\end{equation}
and 
\begin{equation}
    \widetilde{Q}_u(t)~=~\Delta \widetilde{E}_S^{\mathrm{in}}(t)-T_R(0)\Delta S'(t)+T_R(0)\Delta \mathrm{Coh}(t)+T_R(0)\mathcal{C}(t)
\end{equation}
in Eqs. (9) and (10) of the main text. 

From the above expressions, it is evident that only the change of quantum coherence $\Delta \mathrm{Coh}(t)$ during a process plays a role in the bounds. We expect that the sign and thus the contribution of $\Delta \mathrm{Coh}(t)$ to the bounds are process-dependent. Hence we should generally resort to numerical treatments to uncover the detailed role of $\Delta \mathrm{Coh}(t)$ in derived bounds.

In Fig. \ref{fig:erasure_SM}, we show a set of complementary results for the quantum information erasure model (Application 2 of the main text) regarding the role of quantum coherence with varying $\tau$. For the initial state choice with $T_R(0)=T_R$, we remark that both the lower bound $-T\Delta S(t)$ and the derived upper bound $\widetilde{Q}_u(t)+W(t)$ on the dissipated heat contain the same quantum coherence contribution $T_R(0)\Delta\mathrm{Coh}(t)$.
\begin{figure}[thb!]
 \centering
\includegraphics[width=1\columnwidth]{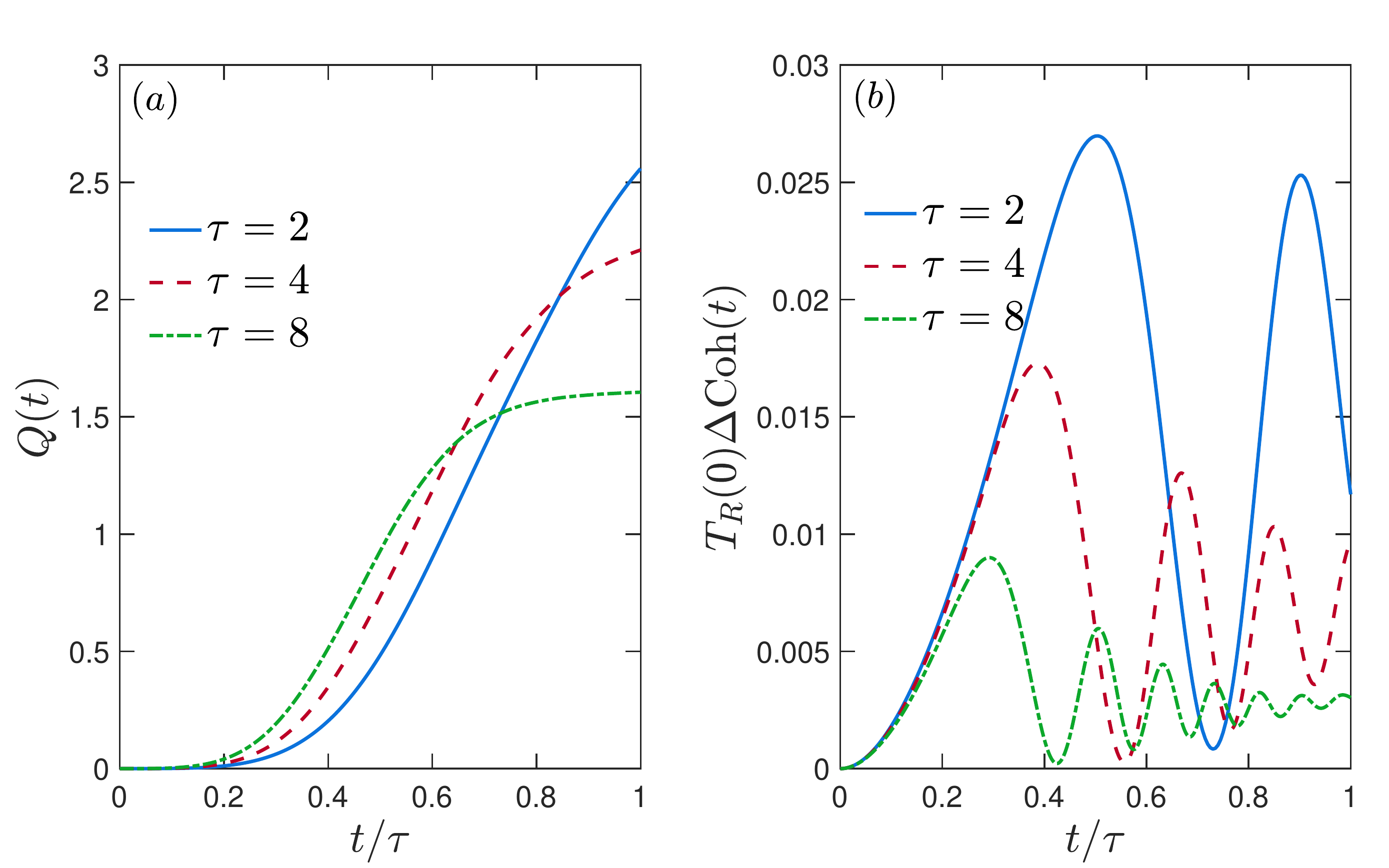} 
\caption{(a) Actual heat dissipation $Q(t)$ with varying $\tau$. (b) Quantum coherence contribution $T_R(0)\Delta\mathrm{Coh}(t)$ with varying $\tau$. Parameters are $\gamma=0.2$, $\beta=T^{-1}=1$, $T_R(0)=1$, $\varepsilon_0=0.4$, $\varepsilon_{\tau}=10$.
}
\protect\label{fig:erasure_SM}
\end{figure}
We note that $\tau$ marks the time span of driving protocols: the smaller the $\tau$ is, the faster the driving fields are. From Fig. \ref{fig:erasure_SM}, the most salient feature is that the oscillation range of the quantum coherence contribution $T_R(0)\Delta\mathrm{Coh}(t)$ tends to be suppressed with increasing $\tau$. This is because faster driving fields will push the system more away from the instantaneous thermal equilibrium state and thus increase the contribution from off-diagonal elements (quantum coherence) of $\rho_S(t)$ in the instantaneous energy basis.

\end{widetext}

\end{document}